\documentclass[12pt]{article}
\usepackage[T1]{fontenc}
\usepackage{helvet}
\usepackage{geometry}
\usepackage{setspace}
\makeatletter

\providecommand{\LyX}{L\kern-.1667em\lower.25em\hbox{Y}\kern-.125emX\@}
\let\SF@@footnote\footnote
\def\footnote{\ifx\protect\@typeset@protect
    \expandafter\SF@@footnote
  \else
    \expandafter\SF@gobble@opt
  \fi
}
\expandafter\def\csname SF@gobble@opt \endcsname{\@ifnextchar[
  \SF@gobble@twobracket
  \@gobble
}
\edef\SF@gobble@opt{\noexpand\protect
  \expandafter\noexpand\csname SF@gobble@opt \endcsname}
\def\SF@gobble@twobracket[#1]#2{}

\newcommand{\lyxaddress}[1]{
  \par {\raggedright #1 
  \vspace{1.4em}
  \noindent\par}
}

\makeatother

\begin{document}

\title{Connecting Green's Functions in an Arbitrary Pair of Gauges and an Application
to Planar Gauges}

\author{Satish D. Joglekar}

\maketitle

\lyxaddress{Department of Physics, I.I.T.Kanpur,Kanpur 208016,INDIA}

\begin{abstract}
We establish a finite field-dependent BRS transformation that connects the Yang
-Mills path-integrals with Faddeev-Popov effective actions for an arbitrary
pair of gauges F and F'.We establish a result that relates an arbitrary Green's
function {[}either a primary one or one that of an operator {]} in an arbitrary
gauge F' to those in gauge F that are \emph{compatible} to the ones in gauge
F \emph{by its construction} {[}in that the construction preserves expectation
values of gauge-invariant observables{]}. We establish parallel results also
for the planar gauge-Lorentz gauge connection.
\end{abstract}

\section{INTRODUCTION}

\paragraph{The importance of Standard Model {[}SM{]} calculations in particle Physics
cannot be overestimated.The Standard Model calculations require the choice of
a gauge.There are a variety of these which have been used in various different
context. Some of these are the Lorentz-type gauges, axial-type gauges,light-cone
gauge,planar gauges,radial gauges,nonlinear gauges , the R\protect\( _{\xi }\protect \)-gauges,Coulomb
gauge etc.Different gauges have been found useful and convenient in different
calculational context {[}1{]}.A priori, we expect from gauge invariance,that
the values for physical observables calculated in different gauges are identical.Formal
proofs of such equivalence for S-matrix elements has been given in a given class
of gauges, say the Lorentz-type gauges with a variable gauge parameter \protect\( \lambda \protect \)
{[}2,3{]}.Some isolated attempts to connect \emph{S-matrix elements} in \emph{singular}
{[}rather than a class of them{]} gauges also have been done.For example, formal
equivalence of S-matrix elements in the Coulomb and the Landau gauges {[}both
singular gauges{]} has also been established {[}2,4{]}.Similar formal attempts
to connect the {[}singular{]} temporal gauge with Feynman gauge in the canonical
formalism has also been done {[}5,6{]}. It is important to note that, however,
the Green's functions in the gauges such as the Coulomb {[}7{]},the axial,the
planar and the light-cone {[}8,9{]} in the path integral formulation are ambiguous
on account of the unphysical singularities in their propagators. Hence, it becomes
important to know how to define the Green's functions in such gauges in such
a manner that they are compatible to those in a well-defined covariant gauge
such as the Lorentz gauge. A general procedure that connects Green's functions
in the path integral formulation in two classes of gauges, say the Lorentz and
the axial,has been lacking until recently. Such comparisons are important not
just in a formal sense but also in practice.Precisely because of this, the proper
treatment of the 1/\protect\( \eta .q\protect \) type poles in axial and light-cone
gauges (and also similar questions in the Coulomb gauge {[}7{]}) has occupied
a lot of attention {[}8,9{]} and the criterion used for their validation has,
in fact, been the comparison with the calculational results in the Lorentz gauges.Such
comparative calculations, where possible {[}8,9 10{]}have to be done by brute
force and have been done toO{[}g\protect\( ^{4}]\protect \) generally; thus
limiting the scope of their confirmation. At a time, a physically observable
anomalous dimension was reported to differ in Lorentz and axial gauges {[}1{]}.
Such questions motivate us to develop a general path integral formalism that
can address all these questions in a wide class of gauges in a single framework.
In a purely Feynman diagrammatic approach,we ,of course, have the attempt of
Cheng and Tsai {[}7{]}.}

\paragraph*{Recently,we developed a general path integral formalism {[}11{]} for connecting
pairs of Yang-Mills effective actions and applied it, in particular, to connecting
the Lorentz and the axial type gauges{[}12,1{]}.{[}In reference 11, we have
also applied the procedure to connecting to the general BRS-anti-BRS invariant
effective action of Baulieu and Thierry-Mieg{]} . Our formalism is based on
the Finite Field-dependent BRS {[}FFBRS{]}transformations {[}11{]} that connect
the two path-integrals.These transformations are of a {[}field-dependent{]}
BRS-type {[}11{]} and are \emph{evaluated in a closed form} and leave the vacuum
expectation value of a gauge-invariant observable \emph{explicitly invariant}
{[}11,13{]} as one transforms from say, a Lorentz-type to an axial-type gauge.
Our procedure, in fact, gives a way of \emph{defining} carefully Green's functions
in the axial-type gauges by a path-integral that explicitly takes care of the
ill-defined nature of the propagator{[}14{]} and in a manner compatible with
those in Lorentz gauges.We found an effective treatment of the axial propagator
using this procedure{[}14,15{]} and applied this formalism also to show the
preservation of the Wilson loop in axial gauges {[}16{]}.To summarize, the output
of the works {[}11-14{]} has been (i) an explicit closed field transformation
in {[}A,c,\protect\( \overline{c}\protect \){]} space to connect the path-integrals
in the two gauges that preserves gauge-invariant observables, (ii)A relation
that allows one to calculate the Green's functions in axial gauges , \emph{compatible
with those in Lorentz gauges by the very construction},(iii)way of dealing with
axial poles that is compatible with the Lorentz gauges. A simplified proof of
(ii) above was also given using the BRS WT-identity {[}17{]}.}

\paragraph{Now that the above techniques have matured,we propose in this work to generalize
our previous works in several directions in this one.In Section 3, we prove
the existence of the FFBRS transformation the connects Faddeev-Popov effective
actions {[}FPEA{]} with any two arbitrary gauge functions F and F'.This applies
to all the gauges mentioned above in the first a paragraph {[} expect the Planar
gauge for which the action in not in the manifest FPEA form{]}.In section 4,
we generalize the identity that connects the Green's functions in gauge F' to
the Green's functions in the gauge F to the case of an arbitrary pair.In section
5, we develop the results for connecting Green's functions in planar gauges
to those in Landau gauge. In section 6, we give a heuristic treatment that uses
the results of section 3 and which connects the vacuum expectation of gauge-invariant
observable in planar gauges to those in Lorentz gauges.This result can also
be verified by explicit calculations that run parallel to those in the section
3;and is not limited,however, by the heuristic treatment we have given for compactness.}

\section{PRELIMINARY}

\paragraph{In this section,we shall review the results in earlier works {[}11-13{]} and
introduce notations.We consider two arbitrary gauge fixing functions F{[}A{]}
and F'{[}A{]} which could be nonlinear or non-covariant and an interpolating
gauge function F\protect\( ^{M}[A]\protect \)= \protect\( \kappa \protect \)F'{[}A{]}+(1-\protect\( \kappa \protect \))F{[}A{]};
~0\protect\( \leq \kappa \leq 1\protect \).~The Faddeev-Popov effective action{[}FPEA{]}
in each case is given by}

\subparagraph{\emph{S\protect\( _{eff}\protect \)} = S\protect\( _{0}\protect \) + S\protect\( _{gf}\protect \)
+\emph{S\protect\( _{gh}\protect \)} ~ ~~ ~~ ~~ ~ ~~ ~~ ~~ ~~~ ~~ ~~ ~~ ~~
~ ~~ ~~ ~~ ~~~ ~~ ~~ ~~ ~ ~~ ~~ ~~ ~~~(2.1)}

with

S\( _{gf} \) = -\( \frac{1}{2\lambda }\int  \) d\( ^{4}x \) \( \widehat{F^{\gamma }} \){[}A{]}\( ^{2} \)
~ ~~ ~~ ~~ ~ ~~ ~~ ~~ ~~~ ~~ ~~ ~~ ~ ~~ ~~ ~~ ~~ ~~ ~~ ~~ ~ ~~ ~~ ~~ ~~~ ~~
~~ ~~~~(2.2a)

and

\emph{S\( _{gh} \)= -}\( \int  \)d\( ^{4}x \) \( \overline{c} \)\( ^{\alpha } \)\( \widehat{M^{\alpha \beta }}c^{\beta } \)
~ ~~ ~~ ~~ ~ ~~ ~~ ~~ ~~~ ~~ ~~ ~~ ~ ~~ ~~ ~~ ~~ ~~ ~~ ~~(2.2b)

with

\( \widehat{M^{\alpha \beta }} \)=\( \frac{\delta }{\delta }\frac{\widehat{F^{\alpha }[A]}}{A^{\gamma }_{\mu }} \)D\( ^{\gamma \beta } \)\( _{\mu } \)
{[}A{]} ~ ~~ ~~ ~~ ~ ~~ ~~ ~~ ~~~ ~~ ~~ ~~ ~ ~~ ~~ ~~ ~~ ~~ ~(2.3)

D\( ^{\alpha \beta } \)\( _{\mu } \) {[}A{]}=\( \delta ^{\alpha \beta }\partial _{\mu }+g_{0}\, f^{\alpha \beta \gamma }A_{\mu }^{\gamma } \)
~ ~~ ~~ ~~ ~ ~~ ~~ ~~ ~~~ ~~ ~~ ~~ ~ ~~ ~~ ~~ ~~ ~~(2.3a)

\paragraph{We denote the FPEA for the three cases \protect\( \widehat{F}\protect \) =F,F',F\protect\( ^{M}\protect \)
by \emph{S\protect\( _{eff}\protect \), S \protect\( '_{eff}\protect \) and
S\protect\( ^{M}\protect \)\protect\( _{eff}\protect \)} respectively.BRS
transformations for the three effective actions are:}

\subparagraph{\protect\( \phi \protect \)'\protect\( _{i}\protect \) =\protect\( \phi \protect \)\protect\( _{i}\protect \)
+\protect\( \delta _{iBRS}[\phi ]\delta \Lambda \protect \)~ ~~ ~~ ~~ ~ ~~
~~ ~~ ~~~ ~~ ~~ ~~ ~~ ~~~ ~~ ~~ ~~ ~ ~~ ~~ ~~ ~~ (2.4)}

\subparagraph{with \protect\( \delta _{iBRS}[\phi ]\protect \) equal to D\protect\( ^{\alpha \beta }\protect \)\protect\( _{\mu }\protect \)c\protect\( ^{\beta }\protect \),-1/2
g\protect\( _{0}f^{\alpha \beta \gamma }\protect \)c\protect\( ^{\beta }\protect \)c\protect\( ^{\gamma }\protect \),and
\protect\( \widetilde{F}/\lambda \protect \) respectively for A,c and \protect\( \overline{c}\protect \).In
the case of the mixed gauge condition, \protect\( \delta _{iBRS}[\phi ]\protect \)
for \protect\( \overline{c}\protect \) is \protect\( \kappa \protect \)-dependent
and in this case we show this explicitly by expressing (2.4) as}

\subparagraph{\protect\( \phi \protect \)'\protect\( _{i}\protect \) =\protect\( \phi \protect \)\protect\( _{i}\protect \)
+\protect\( \widetilde{\delta }_{iBRS}[\phi ,\kappa ]\delta \Lambda \protect \)\protect\( \equiv \phi \protect \)\protect\( _{i}\protect \)
+\protect\( \{\widetilde{\delta _{1}}_{iBRS}[\phi ]+\kappa \protect \)\protect\( \widetilde{\delta _{2}}_{iBRS}[\phi ]\}\delta \Lambda \protect \)~
~~ ~~ ~~ ~~ (2.4a)}

\paragraph{Following observations in {[}11{]}and {[}12{]}, we guess and later prove the
field transformations that takes one from the gauge F to gauge F'.It is given
by the finite field-dependent BRS transformation {[}FFBRS{]} }

\subparagraph{\protect\( \phi \protect \)'\protect\( _{i}\protect \) =\protect\( \phi \protect \)\protect\( _{i}\protect \)
+\protect\( \delta _{iBRS}[\phi ]\Theta \protect \) {[}\protect\( \phi ]\protect \)
~ ~~ ~~ ~~ ~ ~~ ~~ ~~ ~~~ ~~ ~~ ~~ ~ ~~ ~~ ~~ ~~ ~(2.5)}

\subparagraph{where \protect\( \Theta [\phi \protect \){]} has been constructed by the integration
{[}11{]} of the infinitesimal field-dependent BRS {[}IFBRS{]} transformation}

\subparagraph{\protect\( \frac{d\phi _{i}}{d\kappa }=\protect \)\protect\( \delta _{iBRS}[\phi (\kappa )]\Theta \protect \)'{[}\protect\( \phi \protect \)(\protect\( \kappa \protect \)){]}~
~~ ~~ ~~ ~ ~~ ~~ ~~ ~~~ ~~ ~~ ~~ ~ ~~ ~~ ~~ ~~ ~~ ~(2.6)}

(where \( \delta _{iBRS}[\phi (\kappa )] \) refers to the BRS variations of
the gauge F) with the boundary condition \( \phi  \){[}\( \kappa  \)=1{]}=\( \phi  \)'and
\( \phi  \){[}\( \kappa  \)=0{]}=\( \phi  \) and is given in a closed form
by {[}11{]}

\( \Theta  \){[}\( \phi  \){]} = \( \Theta  \)'{[}\( \phi  \){]}{[}exp\{f{[}\( \phi  \){]}\}-1{]}/f{[}\( \phi  \){]}~
~~ ~~ ~~ ~ ~~ ~~ ~~ ~~~ ~~ ~~ ~~ ~ ~~ ~~ ~~ ~~ ~~ ~(2.6a)

and f is given by

f =\( \sum _{i} \) \( \frac{\delta \Theta '}{\delta \phi _{i}} \)\( \delta _{iBRS}[\phi ] \)~~
~~ ~~ ~ ~~ ~~ ~~ ~~~ ~~ ~~ ~~ ~ ~~ ~~ ~~ ~~ ~~ ~(2.6b)

We wish to develop ,in this work, an FFBRS for connecting two arbitrary gauges
F and F'.We shall show that in this case it is given by an FFBRS of the form
(2.5) with \( \Theta  \)'{[}\( \phi  \)(\( \kappa  \)){]} given by

\( \Theta  \)'{[}\( \phi  \)(\( \kappa  \)){]} =\( i\int d^{4}y \) \( \overline{c} \)\( ^{\gamma } \)(y)
(F\( ^{\gamma } \){[}A(\( \kappa )] \) -F' \( ^{\gamma } \){[}A(\( \kappa )] \))
~ ~~ ~~ ~~ ~ ~~ ~~ ~~ ~~ ~~ ~~ ~(2.7)

The Jacobian for the IFBRS transformation is defined as 

D\( \phi  \){[}\( \kappa  \)=0{]} = D\( \phi  \){[}\( \kappa  \){]}J(\( \kappa  \))
= D\( \phi  \){[}\( \kappa  \)+d\( \kappa  \){]}J(\( \kappa  \)+d\( \kappa ) \)
~ ~~ ~~ ~~ ~ ~~ ~~ ~~ ~~~ ~~ ~~ ~~ ~(2.8)

The change in the Jacobian for the IFBRS of (2.6) is given by

-\( \frac{1}{J} \)\( \frac{dJ}{d\kappa }d\kappa  \)=\( \int  \) d\( ^{4}x \)
\( \sum  \)\( _{i} \)(\( \pm ) \)\( \frac{\delta \phi '_{_{i}}(x,\kappa )}{\delta \phi _{i}(y,\kappa )}\Vert _{_{x=y}} \)
~ ~~ ~~ ~~ ~ ~~ ~~ ~~ ~~~ ~~ ~~ ~~ ~ ~~ ~~ ~~ ~~ ~~ ~(2.9)

and is evaluated easily as

-\( \frac{1}{J} \)\( \frac{dJ}{d\kappa }d\kappa  \)=-i\( \int  \)d\( ^{4}x \)\( \{\overline{c} \)(M-M')c
+\( \frac{1}{\lambda }F[F-F'] \)\} ~ ~~ ~~ ~~ ~ ~~ ~~ ~~ ~~~ ~~ ~~ ~~ ~~ ~~
~(2.10)

Further we define

\emph{S\( _{eff} \)\( ^{M}[ \)}\( \phi  \)(\( \kappa  \)),\( \kappa ] \)\( \equiv  \)\emph{S\( _{eff} \)}{[}\( \phi  \)(\( \kappa  \))\emph{\( ] \)+S\( _{1} \)}{[}\( \phi  \)(\( \kappa  \)),\( \kappa ] \)
~ ~~ ~~ ~~ ~ ~~ ~~ ~~ ~~~ ~~ ~~ ~~ ~ ~~ ~~ ~~ ~~ ~~ ~(2.11)

We then have,

\emph{S\( _{1} \)}{[}\( \phi  \)(\( \kappa  \)),\( \kappa ] \)=\( \int  \)d\( ^{4}x \)\{-\( \frac{1}{2\lambda } \){[}\( \kappa ^{2}F'[A(\kappa  \)){]}\( ^{2} \)+
2\( \kappa (1-\kappa )F[A(\kappa  \)){]}\( F'[A(\kappa )] \)

\( +\kappa  \)(\( \kappa -2)F[A(\kappa )]^{2}- \)\( \kappa  \)\( \overline{c}(\kappa )\{ \)M{[}A(\( \kappa )]-M'[A(\kappa )]\}c(\kappa ) \)\}~
~~ ~~ ~~~ ~ ~~ ~~~~~ ~~~ ~~~ ~~(2.12)

We introduce the following notation

<\textcompwordmark{}<f{[}\( \phi  \)(\( \kappa  \)){]}>\textcompwordmark{}>\( _{\kappa } \)\( \equiv  \)
\( \int  \) D\( \phi  \)(\( \kappa  \))f{[}\( \phi  \)(\( \kappa  \)){]}
exp\{ i \emph{S\( _{eff} \)\( ^{M}[ \)}\( \phi  \)(\( \kappa  \)),\( \kappa ] \)\}~
~~ ~~ ~~ ~~~ ~~ ~~ ~~ ~ ~~ ~~ ~~ ~~ ~~ ~(2.13)

In references {[}11{]} and {[}13{]}it was established that the expectation value
of a gauge-invariant observable <\textcompwordmark{}<O{[}A(\( \kappa  \)){]}>\textcompwordmark{}>\( _{\kappa } \)
is independent of \( \kappa  \) iff the Jacobian J(\( \kappa  \)) and the
effective action \emph{S\( _{eff} \)\( ^{M}[ \)}\( \phi  \)(\( \kappa  \)),\( \kappa ] \)
satisfy

<\textcompwordmark{}<\( \frac{i}{J} \)\( \frac{dJ}{d\kappa } \)+\( \frac{dS_{1}[\phi (\kappa ),\kappa ]}{d\kappa } \)>\textcompwordmark{}>\( _{\kappa } \)\( \equiv  \)0
~ ~~ ~~ ~~ ~ ~~ ~~ ~~ ~~~ ~~ ~~ ~~ ~ ~~ ~~ ~~ ~~ ~~ ~(2.14)

In Section 3,we shall verify (2.14) for the IFBRS of (2.6).

\section{FFBRS TRANSFORMATIONS CONNECTING ANY PAIR {[}F,F'{]} OF GAUGES}

\paragraph{In the references {[}11,12{]} explicit field transformations of the FFBRS type
that connected various pairs of equivalent effective actions for the Yang-Mills
theory were constructed.In this work,we wish to give a result that generalizes
it for FPEA in a pair of arbitrary gauges F and F' that includes for example
those mentioned at the beginning of the Introduction {[}except the planar one{]}.These
field transformations are such that they preserve, by an explicit construction,
the expectation values of gauge-invariant observables.}

\paragraph{Consider the expectation value of a gauge invariant observable O{[}A{]} in
the mixed gauge:\footnote{
It is understood that like the Lorentz gauges,an appropriate O(\( \epsilon ) \)
term {[}14,15{]} is necesary in (3.1) to make it well-defined.
}}

\subparagraph{<\textcompwordmark{}<O{[}A{]}>\textcompwordmark{}>\protect\( _{\kappa }\equiv \protect \)
\protect\( \int \protect \) D\protect\( \phi \protect \)(\protect\( \kappa \protect \))O{[}A(\protect\( \kappa \protect \)){]}
exp\{ i \emph{S\protect\( _{eff}\protect \)\protect\( ^{M}[\protect \)}\protect\( \phi \protect \)(\protect\( \kappa \protect \)),\protect\( \kappa ]\protect \)\}
~ ~ ~~ ~~ ~ ~~ ~~ ~~ ~~ ~~ ~(3.1)}

where \( \phi (\kappa ) \) represent fields as defined in (2.6);with the specific
\( \Theta  \)' given by (2.7).

We show that

\( \frac{d}{d\kappa } \)<\textcompwordmark{}<O{[}A{]}>\textcompwordmark{}>\( _{\kappa }\equiv  \)0
~ ~~ ~~ ~~ ~ ~~ ~~ ~~ ~~~ ~~ ~~ ~~ ~ ~~ ~~ ~~ ~~ ~~ ~(3.2)

As shown in {[}13{]}, (3.2) is valid iff

0 \( \equiv  \)\( \int  \) D\( \phi  \)(\( \kappa  \))\{\( \frac{1}{J} \)\( \frac{dJ}{d\kappa } \)-i\( \frac{dS_{1}[\phi (\kappa ),\kappa ]}{d\kappa } \)\}O{[}A(\( \kappa  \)){]}exp\{
i \emph{S\( _{eff} \)\( ^{M}[ \)}\( \phi  \)(\( \kappa  \)),\( \kappa ] \)\}~
~~ ~~ ~~ ~ ~~(3.3)

where we recall from (2.11) {[}noting the definitions below (2.3a){]}

\emph{S\( _{1} \)}{[}\( \phi  \)(\( \kappa  \)),\( \kappa ]= \)\emph{S\( _{eff} \)\( ^{M}[ \)}\( \phi  \)(\( \kappa  \)),\( \kappa ] \)\emph{-S\( ^{}_{eff} \)}{[}\( \phi  \)(\( \kappa  \)){]}
\emph{}~ ~~ ~~ ~~ ~ ~~ ~~ ~~ ~~~ ~~ ~~ ~~ ~ ~~ ~~ ~~ ~~ ~~ ~\emph{(3.4)}

We verify the result (3.3) explicitly below.The proof proceeds much as in {[}11{]}
and {[}12{]}and hence we shall give it briefly.The change in the Jacobian under
the IFBRS of (2.6)

\( \delta  \)\( \phi  \)\( _{i} \)(\( \kappa  \)) =\( \delta _{iBRS}[\phi (\kappa )]\Theta  \)'
{[}\( \phi (\kappa )] \)d\( \kappa  \) 

~ ~~ ~~~ ~~ ~=\( \delta _{iBRS}[\phi (\kappa )] \) \( i\int d^{4}y \) \( \overline{c} \)\( ^{\gamma } \)(y)
(F\( ^{\gamma } \){[}A(\( \kappa )] \) -F' \( ^{\gamma } \){[}A(\( \kappa )] \))d\( \kappa  \)
~~ ~~ ~~ ~~ ~~~ ~~ ~(3.5)

viz.\( \frac{1}{J} \)\( \frac{dJ}{d\kappa }d\kappa  \) is given by (2.10);
viz

\( \frac{1}{J} \)\( \frac{dJ}{d\kappa }d\kappa  \)=i\( \int  \)d\( ^{4}x \)
\{\( \overline{c} \)(M-M')c +\( \frac{1}{\lambda }F[F-F']\} \)~ ~~ ~~ ~~ ~
~~ ~~ ~~ ~~~ ~~ ~~ ~~ ~ ~~ ~~ ~~ ~~ ~~ ~(3.6)

Further,using the expression for \emph{S\( _{1} \)}{[}\( \phi  \)(\( \kappa  \)),\( \kappa ] \)
of (2.12),we find

\( \frac{dS_{1}[\phi (\kappa ),\kappa ]}{d\kappa } \)=\( \underbrace{A} \)\( \Theta  \)'+
\( \underbrace{B} \)~ ~~ ~~ ~~ ~ ~~ ~~ ~~ ~~~ ~~ ~~ ~~ ~ ~~ ~~ ~~ ~~ ~~(3.7)

with

\( \underbrace{A} \)=\( \int  \)d\( ^{4}x \) \( \frac{-1}{\lambda }\{ \)\( \kappa  \)\( ^{2}F'^{\gamma }(M'c)^{\gamma }+\kappa  \)(1-\( \kappa ) \)\( F'^{\gamma }(Mc)^{\gamma }+\kappa (1- \)\( \kappa ) \)\( F^{\gamma }(M'c)^{\gamma }+\kappa ( \)\( \kappa  \)-2)\( F^{\gamma }(Mc)^{\gamma } \)

\( +\kappa  \)\( F^{\gamma }[(M-M')c]^{\gamma }+\kappa  \)\( \int  \)d\( ^{4}y \)d\( ^{4}z \)\( \overline{c} \)\( ^{\gamma }(x)\frac{\delta [F^{\gamma }(x)-F'^{\gamma }(x)]}{\delta A^{\beta }_{\mu }(y)\delta A_{\sigma }^{\eta }(z)} \)(Dc)\( ^{\eta } \)(z)(Dc)\( ^{\beta } \)(y)\}~
~~ ~~ ~ ~~~ ~~ ~~ ~~ ~(3.7a)

and

\( \underbrace{B} \)=\( \int  \)d\( ^{4}x \)\( \{\frac{-1}{\lambda }[\kappa F'^{2}+(1-2\kappa )FF'+(\kappa -1)F^{2}]+ \)
\( \overline{c} \)\( [(M-M')c] \)\} ~ ~~ ~~ ~ ~~~ ~~ ~~ ~~ ~(3.7b)

We note that the last term in \( \underbrace{A} \)vanishes by Bose symmetry;and
\( \underbrace{A} \) can be reorganized further as,

\( \underbrace{A} \)=\( \int  \)d\( ^{4}x \) \( \frac{-1}{\lambda }\{ \)\( \kappa [F'-F][(1-\kappa )Mc+\kappa M'c]\} \)~~
~~ ~ ~~ ~~ ~~ ~~~ ~~ ~~ ~~ ~(3.7c)

The {[}generalized{]} antighost equation of motion is given by

0 \( \equiv  \)\( \int  \) D\( \phi  \)(\( \kappa  \))O{[}A(\( \kappa  \)){]}exp\{
i \emph{S\( _{eff} \)\( ^{M}[ \)}\( \phi  \)(\( \kappa  \)),\( \kappa ] \)\}f(A(\( \kappa  \)),c(\( \kappa  \)),\( \kappa ) \)

~~ ~ ~~ ~~ ~~ ~~~ \( \bullet  \)\{M{[}A(\( \kappa )](1-\kappa )+\kappa M'[A(\kappa )]\}c(\kappa ) \)~
~~ ~ ~~ ~~ ~~ ~~~ ~~ (3.8)

where f{[}A,c,\( \kappa  \){]} is an arbitrary functional of A and c but not
of \( \overline{c} \).

Using (3.8) above,we can convert \( \underbrace{A} \)\( \Theta  \)' term as 

\( \int  \) D\( \phi  \)(\( \kappa  \))O{[}A(\( \kappa  \)){]} \( \underbrace{A} \)\( \Theta  \)'exp\{
i \emph{S\( _{eff} \)\( ^{M}[ \)}\( \phi  \)(\( \kappa  \)),\( \kappa ] \)\}

=-\( \int  \) D\( \phi  \)(\( \kappa  \))O{[}A(\( \kappa  \)){]}\( \Theta  \)'\( \int  \)d\( ^{4}x \)\( \frac{\kappa }{\lambda }[F'-F]^{\gamma } \)\( \frac{1}{i} \)\( \frac{\delta }{\delta \overline{c^{\gamma }(x)}} \)
exp\{ i \emph{S\( _{eff} \)\( ^{M}[ \)}\( \phi  \)(\( \kappa  \)),\( \kappa ] \)\} 

=\( \int  \)d\( ^{4}x \) \( \frac{\kappa }{\lambda }<<[F-F']^{2} \)O{[}A{]}>\textcompwordmark{}>\( _{\kappa } \)
~ ~~ ~~ ~~ ~ ~~ ~~ ~~ ~~~ ~~ ~~ ~~ ~ ~~ ~~ ~~ ~~ ~~ ~(3.9)

Using (3.9),(3.7b) and (3.6), we find

<\textcompwordmark{}<\( \frac{i}{J} \)\( \frac{dJ}{d\kappa } \)+\( \frac{dS_{1}[\phi (\kappa ),\kappa ]}{d\kappa } \)>\textcompwordmark{}>\( _{\kappa } \)\( \equiv  \)0
~ ~~ ~~ ~~ ~ ~~ ~~ ~~ ~~~ ~~ ~~ ~~ ~ ~~ ~~ ~~ ~~ ~~ ~(3.10)

This implies {[} 11,13{]}that <\textcompwordmark{}<O{[}\( \phi  \){]}>\textcompwordmark{}>\( _{\kappa } \)
is independent of \( \kappa  \).Hence,

\( \int  \) D\( \phi  \)O{[}A{]} exp\{ i \emph{S\( _{eff} \)}{[}\( \phi  \){]}\}=\( \int  \)
D\( \phi ' \)O{[}A'{]} exp\{ i \emph{S'\( _{eff} \)}{[}\( \phi ' \){]}\}~~
~~ ~ ~~ ~~ ~~ (3.11)

Thus the FFBRS transformation of (2.5) takes one from gauge F to gauge F' in
the sense already pointed out at the beginning of the section. Such transformations
can be used in establishing a relation between Green's functions in the two
gauges along the lines of {[}13,14{]} that correctly tackle the inherent problems
in many of the gauges.

\section{A RELATION BETWEEN ARBITRARY GREEN'S FUNCTIONS IN TWO GAUGES}

\paragraph{In the work of reference {[}13{]}, we had obtained a relation between arbitrary
Green's functions in axial gauges and Green's functions in the Lorentz gauges
and the former were ``compatible'' with those in the well-defined Lorentz
gauges.This arbitrary Green's function in axial gauges could be expressed either
as {[}i{]} a series of Green's functions in Lorentz gauges that also involve
insertions of the BRS variations in the Lorentz gauges OR {[}ii{]}as an integral
over parameter \protect\( \kappa \protect \) involving the Green's functions
evaluated in the Mixed gauges. It was found from a practical view-point that
the latter result is indeed more amenable to calculations. An axial Green's
function, to a given order, can be evaluated by an integral over \protect\( \kappa \protect \)
of a sum of a finite number of Feynman diagrams in the mixed gauge. This form
has been employed in obtaining an Axial gauge prescription compatible with the
Lorentz gauges {[}15,14{]}. A simpler derivation , based on the BRS, was also
given{[}17{]} for this latter result (only).In this section,we generalize, the
procedure of the reference {[}17{]} to connecting arbitrary Green's functions
in an arbitrary pair of gauges. We emphasize that our treatment in {[}17,13{]}
applies equally well to operator Green's functions {[}needed say in perturbative
QCD applications{]} as to the usual Green's functions.}

\paragraph{In this section,we shall show that the method of Ref.{[}17{]}, based on BRS
WT-identity, can be generalized to a any pair of two gauges F and F'. }

\paragraph{We define\footnote{
As pointed out in section 3,a proper \( \varepsilon -term \) is required to
be added to \emph{S\( _{eff} \)\( ^{M}[ \)}\( \phi  \),\( \kappa ] \) to
make <\textcompwordmark{}<O{[}\( \phi  \){]}>\textcompwordmark{}>\( _{\kappa } \)
well defined.This term depends on F,F' and has to be done seperately in each
specific context.See e.g. the applications in section 5 to planar gauges and
also in {[}14,15{]}.
}, for any operator O{[}\protect\( \phi \protect \){]}, not necessarily local,}

\subparagraph{<\textcompwordmark{}<O{[}\protect\( \phi \protect \){]}>\textcompwordmark{}>\protect\( _{\kappa }\protect \)\protect\( =\int \protect \)
D\protect\( \phi \protect \)O{[}\protect\( \phi \protect \){]}exp\{ i \emph{S\protect\( _{eff}\protect \)\protect\( ^{M}[\protect \)}\protect\( \phi \protect \),\protect\( \kappa ]\protect \)\}~~
~~ ~~ ~ ~~ ~~ ~~ ~~~ ~~ ~ ~~ ~~ ~~ ~(4.1)}

{[}Note:Unlike in section 3 {[}see (3.1){]},we do not now allow the \emph{integration
variable} \( \phi  \) to depend on \( \kappa . \)This suits us here.{]} Then

\( \frac{d}{d\kappa } \)<\textcompwordmark{}<O{[}\( \phi  \){]}>\textcompwordmark{}>\( _{\kappa } \)
=i<\textcompwordmark{}<\( \int  \)d\( ^{4}x \) O{[}\( \phi  \){]}\{-\( \frac{1}{\lambda }[F+\kappa (F'-F)][F'-F]+ \)\( \overline{c} \)(M-M')c>\textcompwordmark{}>\( _{\kappa } \)~~
~ ~~ (4.2)

The WT-identities for \emph{S\( _{eff} \)\( ^{M}[ \)}\( \phi  \),\( \kappa ] \)
are 

<\textcompwordmark{}<\( \int  \)d\( ^{4}x \)\{J\( ^{\alpha } \)\( _{\mu } \)(x)D\( ^{\alpha \beta } \)\( _{\mu } \)c\( ^{\beta } \)+\( \overline{\xi ^{\alpha }} \)\( (x)[-1/2g_{0}f^{\alpha \beta \gamma }c^{\beta }c^{\gamma }](x) \)

~~~~ ~~ ~~ ~~ ~ ~~ ~~ ~~ ~~-\( \xi ^{\alpha }(x) \){[}\( \frac{F+\kappa (F'-F)}{\lambda } \){]}\( ^{\alpha } \)(x)\}>\textcompwordmark{}>
= 0~~ ~ ~ ~~ ~~ ~ (4.3)

Recalling the definition (2.7) of \( \Theta  \)';we operate by

\( \Theta  \)'{[}-i\( \frac{\delta }{\delta J(y)}, \)\( -i\frac{\delta }{\delta \overline{\xi (y)}} \)\( ,i\frac{\delta }{\delta \xi (y)}]\equiv -\int d^{4}y\{F^{\gamma } \){[}-i\( \frac{\delta }{\delta J(y)}]-F'^{\gamma }[ \)-i\( \frac{\delta }{\delta J(y)}]\}\frac{\delta }{\delta \xi ^{\gamma }(y)} \)~
~~ ~~ (4.4)

We then obtain,

<\textcompwordmark{}<\( \int  \)d\( ^{4}x \)\{J\( ^{\alpha } \)\( _{\mu } \)(x)D\( ^{\alpha \beta } \)\( _{\mu } \)c\( ^{\beta } \)+\( \overline{\xi ^{\alpha }} \)\( (x)[-1/2g_{0}f^{\alpha \beta \gamma }c^{\beta }c^{\gamma }](x)-\xi ^{\alpha }(x) \){[}\( \frac{F+\kappa (F'-F)}{\lambda } \){]}\( ^{\alpha } \)(x)\}(-i\( \Theta ') \)

-i\( \int  \)d\( ^{4}x \) (M-M')c(x) \( \overline{c(x)} \) +i\( \int  \)d\( ^{4}x \){[}\( \frac{F+\kappa (F'-F)}{\lambda } \){]}\( ^{\alpha } \){[}F-F'{]}\( ^{\alpha } \)>\textcompwordmark{}>
\( \equiv  \)0~~ ~~ ~~ ~ ~~ ~~ ~~ ~~ (4.5)

Finally, we operate on both sides by -iO{[}-i\( \frac{\delta }{\delta J(y)}, \)\( -i\frac{\delta }{\delta \overline{\xi (y)}} \)\( ,i\frac{\delta }{\delta \xi (y)}] \)
and set J=\( \xi  \)\( =\overline{\xi } \) =0 in the end to obtain

0 = \( \int  \)D\( \phi  \) exp\{ i \emph{S\( _{eff} \)\( ^{M}[ \)}\( \phi  \),\( \kappa ] \)\}\{\( -\int  \){[}\( \frac{\delta O}{\delta A_{\mu }} \)D\( _{\mu }c \)
+\( \frac{\delta ^{R}O}{\delta c^{\alpha }} \)\( [-1/2g_{0}f^{\alpha \beta \gamma }c^{\beta }c^{\gamma }]+ \)\( \frac{\delta ^{R}O}{\delta \overline{c^{\alpha }}} \){[}\( \frac{F+\kappa (F'-F)}{\lambda } \){]}\( ^{\alpha } \){]}(-i\( \Theta ') \)

+O{[}A,c,\( \overline{c} \){]}\{\( \int  \)d\( ^{4}x \)\( \{\overline{c} \)(M-M')c
+\( \frac{F+\kappa (F'-F)}{\lambda }[F-F'] \)\}\}~~ ~~ ~~ ~ ~~ ~~ ~~ ~~~ ~~
~~ ~~ ~ ~~ ~~ ~~ ~~ (4.6)

Using (4.2 ) and (2.4a ) ; we obtain

\( \frac{d}{d\kappa } \)<\textcompwordmark{}<O{[}\( \phi  \){]}>\textcompwordmark{}>\( _{\kappa } \)=
i\( \int  \)D\( \phi  \) exp\{ i \emph{S\( _{eff} \)\( ^{M}[ \)}\( \phi  \),\( \kappa ] \)\}

~ ~ ~~ ~~ ~~ ~~~ ~~ ~~ ~~ ~\( \bullet  \)\{\( \widetilde{\delta }_{iBRS}[\phi ,\kappa ] \)\( \int d^{4}y \)
\( \overline{c} \)\( ^{\gamma } \)(y) (F\( ^{\gamma } \){[}A{]} -F'\( ^{\gamma } \){[}A{]}(y))\( \frac{\delta ^{L}O}{\delta \phi _{i}}\} \)~~
~ ~~ ~~~ ~ ~~ ~~~ (4.7)

Integrating from \( \kappa  \) = 0 to 1 we obtain

<\textcompwordmark{}<O{[}\( \phi  \){]}>\textcompwordmark{}>\( _{F} \)'= <\textcompwordmark{}<O{[}\( \phi  \){]}>\textcompwordmark{}>\( _{F} \)
+i\( \int  \)\( _{_{0}} \)\( ^{1} \)d\( \kappa  \) D\( \phi  \) exp\{ i
\emph{S\( _{eff} \)\( ^{M}[ \)}\( \phi  \),\( \kappa ] \)\}

~ ~~ ~~ ~~ ~~~ ~~ ~ ~~ ~~ ~~ ~~~ ~~~~ ~~ ~\( \bullet  \)\( \sum _{i}(\widetilde{\delta _{1i}[\phi } \){]}+\( \kappa  \)\( (\widetilde{\delta _{2i}[\phi } \){]})(-i\( \Theta ') \)\( \frac{\delta ^{L}O}{\delta \phi _{i}} \)~~
~ ~~~ ~~ ~ ~~ ~~ ~~ ~~~ ~~~~ (4.8)

This gives us the expression for the Green's function {[}which depending on
the choice of O{[}\( \phi  \){]} could be a primary one or of an operator{]}
in one gauge F' related to that in F ,with finite number of additional terms
that can be evaluated by Feynman diagram techniques as mentioned in the beginning
of this section.

\section{Planer Gauges}

\paragraph{Planar gauges have been found very useful in the perturbative QCD calculations
on account of a simpler propagator structure {[}that avoids the double pole
in (\protect\( \eta \protect \).k){]} and other attractive features {[}18{]}.
Gauges similar to the planer gauges have also been used in the renormalization
{[}19{]}of higher derivative gravity theories.}

\paragraph{Planar gauges are defined by\footnote{
For mathematical rigor,we may replace \( \partial ^{2} \)by \( \partial ^{2}-i\epsilon  \)
in the equations (5.1),(5.2) and (5.5) and in the definition of \( \overline{c} \)''
in (5.3).
}}

\subparagraph{S\protect\( ^{P}\protect \)\protect\( _{gf}\protect \) = -\protect\( \frac{1}{2\lambda \eta ^{2}}\int \protect \)
d\protect\( ^{4}x\protect \) \protect\( \eta \protect \).A \protect\( \partial ^{2}\protect \)\protect\( \eta \protect \).A
;~ ~~ ~~ ~ \protect\( \eta \protect \)\protect\( ^{2}\protect \)\protect\( \neq \protect \)0~,\protect\( \lambda =1.\protect \)
~ ~~ ~~ ~ ~~~ ~~ (5.1)}

\subparagraph{and the accompanying ghost term}

\subparagraph{\emph{S\protect\( ^{P}\protect \)\protect\( _{gh}\protect \)} =\emph{-}\protect\( \int \protect \)d\protect\( ^{4}x\protect \)
\protect\( \overline{c}\protect \)\protect\( ^{\alpha }\protect \)\protect\( \partial ^{2}\protect \)\protect\( \eta \protect \)\protect\( .Dc^{\alpha }\protect \)~~
~~ ~~ ~ ~~ ~~ ~~ ~~~ ~~ ~~ (5.2)}

or equivalently

\emph{S\( ^{P} \)\( _{gh} \)} \emph{\( \equiv  \)-}\( \int  \)d\( ^{4}x \)
\( \overline{c} \)''\( ^{\alpha } \)\( \eta  \)\( .Dc^{\alpha } \)~~ ~~
~~ ~ ~~ ~~ ~~ ~~~ ~~ ~~ ~~ ~ ~~ ~~ ~~ ~~ (5.3)

with \( \overline{c} \)''\( ^{\alpha } \) defined as \( \partial ^{2} \)\( \overline{c} \)\( ^{\alpha } \).The
net FPEA for the planar gauges,S\( ^{P} \)\( _{eff} \),has a BRS invariance
under 

\( \delta \overline{c} \) =\( \frac{\eta .A}{\eta ^{2}\lambda }\delta \Lambda  \)~~
~~ ~~ ~ ~~ ~~ ~~ ~~~ ~~ ~~ ~~ ~ ~~ ~~ ~~ ~~ ~~ ~~~ ~~ ~~ ~~ ~ ~~ ~~ ~~ ~~ (5.4)

and \( \delta A \) and \( \delta c \) as in (2.4 )

\paragraph{We note the the gauge fixing term is not manifestly of the form of (2.2) viz.
\~{} \protect\( \int \protect \)d\protect\( ^{4}x\protect \) \protect\( \widehat{F^{\gamma }}\protect \){[}A{]}\protect\( ^{2}\protect \).Hence,we
cannot apply the results established in the last two sections directly to this
case \footnote{
See, however, an alternate way presented in Section 6.
}.We can establish a route to connect this gauge to the Lorentz gauge as follows:}

\subparagraph{S\protect\( ^{P}\protect \)\protect\( _{gf}\protect \) --> S\protect\( ^{\widetilde{L}}\protect \)\protect\( _{gf}\protect \)
\protect\( \equiv \protect \)-\protect\( \frac{1}{2\lambda \eta ^{2}}\int \protect \)
d\protect\( ^{4}x\protect \) \protect\( \partial \protect \).A \protect\( \partial ^{2}\protect \)\protect\( \partial \protect \).A
-->S\protect\( ^{L}\protect \)\protect\( _{gf}\protect \)~~ ~~ ~~ ~ ~~ ~~
~~ ~~~ ~~ (5.5a)}

\subparagraph{or as,}

\subparagraph{S\protect\( ^{P}\protect \)\protect\( _{gf}\protect \) --> S\protect\( ^{A}\protect \)\protect\( _{gf}\protect \)
-->S\protect\( ^{L}\protect \)\protect\( _{gf}\protect \)~ ~~ ~~ ~ ~~ ~~ ~~
~~~ ~~ ~~ ~~ ~ ~~ ~~ ~~ ~~ (5.5b)}

\paragraph*{We shall call, for the sake of nomenclature,the gauge \protect\( \widetilde{L}\protect \)
the ``pseudo-Lorentz'' gauge.The formal connection between the pseudo-Lorentz
and the Lorentz gauges as well as that between planar and the axial gauge can
be established easily from the work of Lee and Zinn-Justin {[}2{]} itself.We
note from {[}2{]},}

\paragraph*{<\textcompwordmark{}<O{[}A{]}>\textcompwordmark{}>\protect\( _{f}\protect \)\protect\( \equiv \protect \)\protect\( \int \protect \)
D\protect\( \phi \protect \) exp\{ i \emph{S}\protect\( _{0}+iS^{A}_{G}\protect \)\}\protect\( \begin{array}{c}
\Pi \\
\alpha ,x
\end{array}\delta \protect \) (\protect\( \eta \protect \).A\protect\( ^{\alpha }-f^{\alpha }\protect \)
)O{[}A{]} ~ ~~ ~~ (5.6)}

\paragraph*{is independent of 'f' for any gauge-invariant observable O{[}A{]}.Now, the
vacuum expectation values of gauge-invariant observable O{[}A{]} in the planar
and the axial gauges are related to the above quantity <\textcompwordmark{}<O{[}A{]}>\textcompwordmark{}>\protect\( _{f}\protect \)
by the relations:}

\subparagraph{<\textcompwordmark{}<O{[}A{]}>\textcompwordmark{}>\protect\( _{A}\protect \)\protect\( =\protect \)\protect\( \int \protect \)
Df exp\{\protect\( -\frac{i}{2\lambda }\protect \)\protect\( \int \protect \)
d\protect\( ^{4}x\protect \)f\protect\( ^{2}\protect \)\} <\textcompwordmark{}<O{[}A{]}>\textcompwordmark{}>\protect\( _{f}\protect \)~~
~~ ~~ ~~~ ~~ ~~ ~~ ~ ~~ ~~ ~~ ~~ (5.7)}

\subparagraph{and\footnote{
For mathematical rigor,we may replace \( \partial ^{2} \)by \( \partial ^{2}-i\epsilon  \)
,as earlier, in the equation below and everywhere else necessary.
}}

\subparagraph{<\textcompwordmark{}<O{[}A{]}>\textcompwordmark{}>\protect\( _{P}\protect \)=\protect\( \frac{1}{N_{P}}\protect \)\protect\( \int \protect \)
Df exp\{\protect\( -\frac{i}{2\lambda \eta ^{2}}\protect \)\protect\( \int \protect \)
d\protect\( ^{4}x\protect \)f \protect\( \partial ^{2}\protect \)f\}<\textcompwordmark{}<O{[}A{]}>\textcompwordmark{}>\protect\( _{f}\protect \)~~
~~ ~~~ ~~ ~~ ~~ ~ ~~ ~~ ~~ ~~ (5.8)}

\subparagraph{{[}Equation (5.8) is compensated by appropriate normalization factor \protect\( \frac{1}{N_{P}}\protect \)\emph{relative
to} (5.7){]}.Therefore, in view of the f-independence of <\textcompwordmark{}<O{[}A{]}>\textcompwordmark{}>\protect\( _{f}\protect \),
the two quantities above are equal:}

\subparagraph{<\textcompwordmark{}<O{[}A{]}>\textcompwordmark{}>\protect\( _{A}=\protect \)<\textcompwordmark{}<O{[}A{]}>\textcompwordmark{}>\protect\( _{P}\protect \)~
~~ ~~~ ~~ ~~ ~~ ~ ~~ ~~~ ~~ ~~ ~~ ~ ~~~ ~~ ~~ ~~ ~~ (5.9)}

\subparagraph{Thus, as for as the gauge-invariant observables are concerned, once the Lorentz
-axial route is established, Lorentz--planar gauge connection also becomes available.We
shall establish this explicit connection later in section 6using the results
of Sections 3. But first, we would like to derive a result similar to that in
Section 4 for the the planar--Lorentz Green's functions connection applicable
to the \emph{arbitrary} Green's functions.}

We define

W\( ^{P}[J,\xi ,\overline{\xi }]= \)\( \frac{1}{N_{P}} \)\( \int  \) D\( \phi  \)
exp\{ i \emph{S}\( _{0}+iS^{A}_{G}-\frac{i}{2\lambda \eta ^{2}} \)\( \int  \)
d\( ^{4}x \) \( \eta  \).A \( \partial ^{2} \)\( \eta  \).A+ Source Terms\}~~
(5.10)

As for the ghost term,we can use an identical one both for the planar and the
axial gauges ,obtained from (5.2) by a field redefinition as in (5.3).

S\( ^{A}_{G}= \)\emph{-}\( \int  \)d\( ^{4}x \) \( \overline{c} \)\( ^{\alpha } \)\( \eta  \)\( .Dc^{\alpha } \)
~~ ~~ ~~ ~ ~~ ~~ ~~ ~~~ ~~ ~~ ~~ ~ ~~ ~~ ~~ ~~ (5.11)

We now define, for the singular axial gauge (\( \eta  \).A\( ^{\alpha }-f^{\alpha } \)
)=0,

W\( _{f} \)\( ^{}[J,\xi ,\overline{\xi }] \)\( \equiv  \)\( \int  \) D\( \phi  \)
exp\{ i \emph{S}\( _{0}+iS^{A}_{G} \)+ Source Terms\}\( \begin{array}{c}
\Pi \\
\alpha ,x
\end{array}\delta  \) (\( \eta  \).A\( ^{\alpha }-f^{\alpha } \) )~ ~~ ~~ (5.12)

We then have 

W\( ^{P}[J,\xi ,\overline{\xi }]= \)\( \frac{1}{N_{P}} \)\( \int  \) Df exp\{\( -\frac{i}{2\lambda \eta ^{2}} \)\( \int  \)
d\( ^{4}x \)f \( \partial ^{2} \)f\}W\( _{f} \) \( [J,\xi ,\overline{\xi }] \)~~
~~ ~~ ~ ~~ ~~ ~~ ~~~ ~~ ~~ ~~ ~ ~~ ~~ ~~ ~~ (5.13)

W\( ^{A}[J,\xi ,\overline{\xi }]= \)\( \int  \) Df exp\{\( -\frac{i}{2\lambda } \)\( \int  \)
d\( ^{4}x \)f\( ^{2} \)\}W\( _{f} \)\( ^{}[J,\xi ,\overline{\xi }] \)~~
~~ ~~ ~ ~~ ~~ ~~ ~~~ ~~ ~~ ~~ ~ ~~ ~~ ~~ ~~ (5.14)

We already know how the Green's functions in Axial gauges are linked to those
in Lorentz gauges{[}13,17{]}.To relate the Green's functions in the planar gauge
to those in Lorentz gauges, we proceed as follows:

Consider the expectation value of an arbitrary operator O{[}\( \phi  \){]},
possibly multi-local,in the Lorentz gauge, the singular axial gauge (\( \eta  \).A\( ^{\alpha }-f^{\alpha } \)
)=0 and in the planar gauge:

<\textcompwordmark{}<O{[}\( \phi  \){]}>\textcompwordmark{}>\( _{L}= \)\( \int  \)D\( \phi  \)O{[}\( \phi  \){]}exp\{
i \emph{S\( _{eff} \)\( ^{L}[ \)}\( \phi  \){]}+ \( \varepsilon O_{L} \)\}~~~
~~ ~~ ~~ ~ ~~ ~~ ~~ ~~ (5.15)

where \( \varepsilon O_{L} \) are the \( \varepsilon  \)-terms in the Lorentz
gauges:

\( \varepsilon O_{L} \) = \( \int  \) d\( ^{4}x \) {[}\( \frac{1}{2} \)A\( _{\mu } \)A\( ^{\mu } \)
-\( \overline{c} \)\( ^{} \)c {]}~ ~~ ~~ ~~ ~ ~~ ~~ ~~~ ~~ ~~ ~ ~~ ~~ ~~ ~~
(5.16)

<\textcompwordmark{}<O{[}\( \phi  \){]}>\textcompwordmark{}>\( _{f}= \)\( \int  \)D\( \phi  \)O{[}\( \phi  \){]}exp\{
i \emph{S\( _{0} \)}{[}\( \phi  \){]}+iS\( ^{_{A}}_{G}[\phi  \){]}+ \( \varepsilon O_{A}(f) \)\}\( \begin{array}{c}
\Pi \\
\alpha ,x
\end{array}\delta  \) (\( \eta  \).A\( ^{\alpha }-f^{\alpha } \) )~~ ~~~ (5.17)

where \( \varepsilon O_{A}(f) \) terms are the appropriate O{[}\( \varepsilon ] \)
terms arrived at as in {[}14{]} that can depend on f;and

<\textcompwordmark{}<O{[}\( \phi  \){]}>\textcompwordmark{}>\( _{P} \)=\( \int  \)D\( \phi  \)O{[}\( \phi  \){]}exp\{
i \emph{S\( _{eff} \)\( ^{P}[ \)}\( \phi  \){]}+ \( \varepsilon O_{P} \)\}~~
~~~ ~~ ~~ ~ ~~ ~~ ~~ ~~ (5.18)

From (5.18 ) and (5.17),

<\textcompwordmark{}<O{[}\( \phi  \){]}>\textcompwordmark{}>\( _{P} \) \( \equiv  \)\( \frac{1}{N_{P}} \)\( \int  \)
Df exp\{\( -\frac{i}{2\lambda \eta ^{2}} \)\( \int  \) d\( ^{4}x \)f \( \partial ^{2} \)f\}<\textcompwordmark{}<O{[}\( \phi  \){]}>\textcompwordmark{}>\( _{f} \)~
~~~ ~~ ~~ ~ ~~ ~~ ~~ ~~ (5.19)

{[}and the above relation in fact determines what \( \varepsilon O_{P} \) should
be{]}.We note\footnote{
For mathematical rigor,we may replace \( \sigma  \) \( \rightarrow  \)\( \frac{\sigma }{1-i\epsilon '} \)
with \( \epsilon ' \)> 0 in what follows.We have used:\( \begin{array}{c}
lim\\
\sigma \rightarrow 0
\end{array} \)\( \sqrt{\frac{i+\epsilon '}{2\pi \sigma ^{2}}} \)exp {[}\( -\frac{ix^{2}[1-i\epsilon ']}{\sigma ^{2}}] \)
= \( \delta  \) (x).
} ,

<\textcompwordmark{}<O{[}\( \phi  \){]}>\textcompwordmark{}>\( _{f} \) =\( \begin{array}{c}
lim\\
\sigma \rightarrow 0
\end{array} \)\( \int  \)D\( \phi  \)O{[}\( \phi  \){]}exp\{ i \emph{S\( _{0} \)}{[}\( \phi  \){]}+iS\( ^{_{A}}_{G}[\phi  \){]}-\( \frac{i}{2\sigma }\int  \)
d\( ^{4}x \) (\( \eta  \).A-f)\( ^{2} \) + \( \varepsilon O_{A}(f,\sigma ) \)\}

~~~ ~~ ~~ ~~ ~~ ~~ ~=\( \begin{array}{c}
lim\\
\sigma \rightarrow 0
\end{array} \)<\textcompwordmark{}<O{[}\( \phi  \){]}>\textcompwordmark{}>\( _{f} \)\( _{,\sigma } \)~~~
~~ ~ (5.20)

Now, <\textcompwordmark{}<O{[}\( \phi  \){]}>\textcompwordmark{}>\( _{f} \)\(  \)\( _{,\sigma } \)
can be related to <\textcompwordmark{}<O{[}\( \phi  \){]}>\textcompwordmark{}>\( _{L,} \)\( _{,\sigma } \)
by the result (4.8).We thus have,

<\textcompwordmark{}<O{[}\( \phi  \){]}>\textcompwordmark{}>\( _{f} \)\( _{,\sigma } \)
=<\textcompwordmark{}<O{[}\( \phi  \){]}>\textcompwordmark{}>\( _{L,} \)\( _{\sigma } \) 

~ ~~ ~~ ~ ~~ ~ ~~+i\( \frac{1}{N_{P}} \)\( \int  \)\( _{_{0}} \)\( ^{1} \)d\( \kappa  \)\( \int  \)D\( \phi  \)
exp\{iS\( _{0}+ \) i \emph{S\( _{G} \)\( ^{M}[ \)}\( \phi  \),\( \kappa ] \)
-\( \frac{i}{2\sigma }\int  \) d\( ^{4}x \) F\( _{1}^{M} \){[}A{]}\( ^{2} \)+\( \varepsilon O_{L} \)\}

~~ ~~ ~~ ~ ~~ ~ ~~ ~~ ~~ ~~ ~\( \bullet  \)\( \sum _{i}(\widetilde{\delta _{1i}[\phi } \){]}+\( \kappa  \)\( (\widetilde{\delta _{2i}[\phi } \){]})(-i\( \Theta '_{_{1}}) \)\( \frac{\delta ^{L}O}{\delta \phi _{i}} \)~~
~~ ~ ~~ ~~ ~~ ~~ (5.21)

Here,on account of the results in Section 4,we have 

F\( _{1}^{M} \){[}A{]}=\( \partial .A(1-\kappa )+\kappa (\eta .A-f) \) ~~
~~ ~ ~~ ~~ ~~ ~ ~~ ~ ~~ ~~ ~~ ~~ ~~ ~ ~~ ~~ ~~ ~~ (5.22)

and

\( \Theta '_{_{1}}= \)\( i\int d^{4}y \) \( \overline{c} \)\( ^{\gamma } \)(y)
(\( \partial  \).A\( ^{\gamma } \)-\( \eta  \).A\( '^{\gamma } \)+f\( ^{\gamma } \))~
~~ ~ ~~ ~~ ~~ ~~ ~~ ~ ~~ ~~ ~~ ~~ (5.23)

\emph{S\( _{G} \)\( ^{M}[ \)}\( \phi  \),\( \kappa ] \)=-\( \int  \) d\( ^{4}x \)\( \overline{c} \)
{[}M\( (1-\kappa )+\kappa M']c \)~ ~~ ~~ ~~ ~~ ~~ ~ ~~ ~~ ~~ ~~ (5.24)

We substitute (5.21) in (5.19) to obtain,

<\textcompwordmark{}<O{[}\( \phi  \){]}>\textcompwordmark{}>\( _{P} \) \( \equiv  \)\( \begin{array}{c}
lim\\
\sigma \rightarrow 0
\end{array} \)\( \frac{1}{N_{P}} \)\( \int  \) Df exp\{\( -\frac{i}{2\lambda \eta ^{2}} \)\( \int  \)
f \( \partial ^{2} \)f\}\{<\textcompwordmark{}<O{[}\( \phi  \){]}>\textcompwordmark{}>\( _{L,} \)\( _{,\sigma } \)
+i\( \int  \)\( _{_{0}} \)\( ^{1} \)d\( \kappa  \) \( \int  \)D\( \phi  \)
exp\{iS\( _{0}+ \) i \emph{S\( _{G} \)\( ^{M}[ \)}\( \phi  \),\( \kappa ] \)
-\( \frac{i}{2\sigma }\int  \) d\( ^{4}x \) F\( _{1}^{M} \){[}A{]}\( ^{2} \)+\( \varepsilon O_{L} \)\( \mid  \)\}\( \sum _{i}(\widetilde{\delta _{1i}[\phi } \){]}+\( \kappa  \)\( (\widetilde{\delta _{2i}[\phi } \){]})(-i\( \Theta '_{_{1}}) \)\( \frac{\delta ^{L}O}{\delta \phi _{i}} \)~~
~~ ~ ~~ ~~ ~~ ~~ (5.25)

As <\textcompwordmark{}<O{[}\( \phi  \){]}>\textcompwordmark{}>\( _{L,} \)\( _{,\sigma } \)
is independent of 'f' and \( \int  \) Df exp\{\( -\frac{i}{2\lambda \eta ^{2}} \)\( \int  \)
d\( ^{4}x \)f \( \partial ^{2} \)f\} is absorbed in normalization,we arrive
at

<\textcompwordmark{}<O{[}\( \phi  \){]}>\textcompwordmark{}>\( _{P} \) = <\textcompwordmark{}<O{[}\( \phi  \){]}>\textcompwordmark{}>\( _{L,} \)\( _{\sigma } \)\( _{\rightarrow } \)
\( _{0} \) +\( \begin{array}{c}
lim\\
\sigma \rightarrow 0
\end{array} \)\( \int  \)\( _{_{0}} \)\( ^{1} \)d\( \kappa  \) I(\( \sigma ,\kappa )d\kappa  \)~~
~~ ~~ ~~ ~ ~~ ~~ ~~ ~~ (5.26)

where

I(\( \sigma ,\kappa )=i \)\( \frac{1}{N_{P}} \)\( \int  \)Df exp\{\( -\frac{i}{2\lambda \eta ^{2}} \)\( \int  \)d\( ^{4}x \)f\( \partial ^{2} \)f\}\( \int  \)D\( \phi  \)
exp\{iS\( _{0}+ \)i\emph{S\( _{G} \)\( ^{M}[ \)}\( \phi  \),\( \kappa ] \)-\( \frac{i}{2\sigma }\int  \)d\( ^{4}x \)
{[}\( \partial .A(1-\kappa )+\kappa (\eta .A-f) \){]}\( ^{2} \)+\( \varepsilon O_{L} \)\( \mid  \)\}\( \sum _{i}(\widetilde{\delta _{1i}[\phi } \){]}+\( \kappa  \)\( (\widetilde{\delta _{2i}[\phi } \){]})(-i\( \Theta '_{_{1}}) \)\( \frac{\delta ^{L}O}{\delta \phi _{i}} \)~~
~~ ~ ~~ ~~ ~~ ~~ (5.27)

Now, we take the limit \( \sigma \rightarrow  \)0 inside to obtain

exp\{-\( \frac{i}{2\sigma }\int  \) d\( ^{4}x \) F\( _{1}^{M} \){[}A{]}\( ^{2} \)\}\( \sim  \)
\( \begin{array}{c}
\Pi \\
\alpha ,x
\end{array} \)\( \delta  \)\{\( \partial .A(1-\kappa )+\kappa (\eta .A-f) \)\}~~ ~~ ~~ ~
~~ ~~ ~ ~~ ~~ ~~ ~~ (5.28)

We further note that

D\( \overline{c} \)\( \begin{array}{c}
\Pi \\
\alpha ,x
\end{array} \)\( \delta  \)\{\( \partial .A(1-\kappa )+\kappa (\eta .A-f) \)\}\( \equiv  \)\( \begin{array}{c}
\Pi \\
\alpha ,x
\end{array} \)d\( \overline{c} \)\( ^{\alpha } \)(x)\( \delta  \)\{\( \partial .A(1-\kappa )+\kappa (\eta .A-f) \)\}

= \( \begin{array}{c}
\Pi \\
\alpha ,x
\end{array} \)d\( \underbrace{\overline{c}^{\alpha }} \)(x)\( \delta  \)\{\( \partial .A(1/\kappa -1)+(\eta .A-f) \)\}

\( \sim  \)D\( \underbrace{\overline{c}} \)\( \begin{array}{c}
\Pi \\
\alpha ,x
\end{array} \)\( \delta  \)\{\( \partial .A(1/\kappa -1)+(\eta .A-f) \)\}~ ~~ ~~ ~~ ~ ~~~~
~ ~~~ ~~ ~ ~~ ~~ (5.29)

with \( \underbrace{\overline{c}} \)=\( \overline{c} \)\( \kappa  \). We
use this \( \delta  \)-function to simplify

\(  \)\( \Theta '_{_{1}}= \)\( i\int d^{4}y \) \( \overline{c} \)\( ^{\gamma } \)(y)
(\( \partial  \)\( _{^{..A^{\gamma }}} \))/\( \kappa  \) =~ ~ \( i\int d^{4}y \)
\( \underbrace{\overline{c}} \)\( ^{\gamma } \)(y) (\( \partial  \).A\( ^{\gamma } \))/\( \kappa  \)\( ^{2} \)~~
~~ ~~ ~~ ~~ ~ ~~ ~~ ~~ ~~ (5.30)

~and 

exp\{\( -\frac{i}{2\lambda \eta ^{2}} \)\( \int  \)d\( ^{4}x \)f\( \partial ^{2} \)f\}-->exp\{\( -\frac{i}{2\lambda \eta ^{2}} \)\( \int  \)d\( ^{4}x \){[}\( \partial .A(1/\kappa -1)+\eta .A \){]}\( \partial ^{2} \){[}\( \partial .A(1/\kappa -1)+\eta .A \){]}

~~ ~~ ~~ ~~ ~ ~~ ~~ \( \equiv  \)exp\{i\( \underbrace{S} \)\( _{gf} \)(\( \kappa )\} \)~
~ ~~ (5.31)

We further re-express \emph{S\( _{G} \)\( ^{M} \) as}

\emph{S\( _{G} \)\( ^{M} \)=}-\( \int  \) d\( ^{4}x \) \( \underbrace{\overline{c}} \){[}M(1/\( \kappa  \)-1)\( +M']c \)~\( \equiv  \)
\( \underbrace{S} \)\emph{\( _{G} \)\( ^{M} \)}~~ ~~ ~~ ~~ ~~ ~ ~~ ~~ ~~
~~ (5.32)

Thus, we obtain the result that connects arbitrary Green's functions in planar
gauges to those in Landau gauge:

<\textcompwordmark{}<O{[}\( \phi  \){]}>\textcompwordmark{}>\( _{P} \) =<\textcompwordmark{}<O{[}\( \phi  \){]}>\textcompwordmark{}>\( _{L,} \)\( _{\sigma } \)\( _{\rightarrow } \)\( _{0} \)+

i\( \frac{1}{N_{P}} \)\( \int  \)\( _{_{0}} \)\( ^{1} \)d\( \kappa  \)\( \int  \)D\( \underbrace{\phi } \)exp\{iS\( _{0}+ \)i\( \underbrace{S} \)\emph{\( _{G} \)\( ^{M}[ \)}\( \phi  \),\( \kappa ] \)+i\( \underbrace{S} \)\( _{gf} \)(\( \kappa )\} \)+\( \varepsilon O_{L} \)\}

~ ~~ ~~ ~~ ~~ ~ ~~ ~~ ~~ ~ ~~ ~~ ~~ ~~ ~ ~~ ~~ ~~\( \bullet  \)\( \sum _{i}(\widetilde{\delta _{1i}[\phi } \){]}+\( \kappa  \)\( (\widetilde{\delta _{2i}[\phi } \){]})(-i\( \Theta '_{_{1}}) \)\( \frac{\delta ^{L}O}{\delta \phi _{i}} \)~~
~~ ~ ~~ ~~ ~~ (5.33)

We may change the notation for the integration variable \( \underbrace{\overline{c}} \)
\( \rightarrow  \) \( \overline{c} \).Noting that

\( \sum _{i}(\widetilde{\delta _{1i}[\phi } \){]}+\( \kappa  \)\( (\widetilde{\delta _{2i}[\phi } \){]}
does not depend on this variable,we can express this as

<\textcompwordmark{}<O{[}\( \phi  \){]}>\textcompwordmark{}>\( _{P} \) =<\textcompwordmark{}<O{[}\( \phi  \){]}>\textcompwordmark{}>\( _{landau} \)+i\( \frac{1}{N_{P}} \)\( \int  \)\( _{_{0}} \)\( ^{1} \)\( \frac{d\kappa }{\kappa ^{2}} \)
\( \int  \)D\( \phi  \) exp\{i\( \underbrace{S} \)\( ^{M}_{eff} \)+\( \varepsilon [\frac{1}{2} \)A\( _{\mu } \)A\( ^{\mu } \)-
i

\( \overline{c} \)c {]}\}\( \sum _{i}(\widetilde{\delta _{1i}[\phi } \){]}+\( \kappa  \)\( (\widetilde{\delta _{2i}[\phi } \){]}){[}\( \int d^{4}y \)
\( \overline{c} \)\( ^{\gamma } \)(y) (\( \partial  \).A\( ^{\gamma } \)){]}\( \frac{\delta ^{L}O[A,c,\kappa \overline{c}]}{\delta \underbrace{\phi _{i}}} \)~~
~~ ~ ~~ ~~ ~~ ~~ (5.34)

with \( \underbrace{\phi _{i}} \)=A,c,\( \kappa \overline{c} \);and here

\( \underbrace{S} \)\( ^{M}_{eff} \) = S\( _{0} \)\( -\frac{1}{2\lambda \eta ^{2}} \)\( \int  \)
d\( ^{4}x \){[}\( \partial .A(1/\kappa -1)+\eta .A \){]}\( \partial ^{2} \){[}\( \partial .A(1/\kappa -1)+\eta .A \){]}-\( \int  \)
d\( ^{4}x \) \( \underbrace{\overline{c}} \){[}M(1/\( \kappa  \)-1)\( +M']c \)~
~~ ~~ ~~ ~~ (5.35)

is the effective action for a mixed planar gauge but with a mixed gauge fixing
function: {[}\( \partial .A(1/\kappa -1)+\eta .A \){]} in both the gauge-fixing
and the ghost term.We put O{[}\( \phi  \){]}=I, the identity operator, in (5.34)
to obtain,

<\textcompwordmark{}< I >\textcompwordmark{}>\( _{P} \) =<\textcompwordmark{}<
I >\textcompwordmark{}>\( _{landau} \)~~ ~ ~~ ~~ ~~ ~ ~ ~~ ~~ ~~ ~~ ~ ~ ~~
~~ ~~ ~~~ ~~ ~ ~ ~~ ~~ ~~ ~~ ~ ~ ~~ ~~ ~~ ~~ (5.36)

We then obtain for the connected part <O{[}\( \phi  \){]}>\( _{P} \) =<\textcompwordmark{}<O{[}\( \phi  \){]}>\textcompwordmark{}>\( _{P} \)/<\textcompwordmark{}<
I >\textcompwordmark{}>\( _{P} \) ,etc the final result\footnote{
We need not be alarmed by \( \kappa ^{2} \) in the denominator : near \( \kappa  \)=0,
the ghost propagator and the longitudinal gauge propagator yield enough factors
of \( \kappa  \).
}:

<O{[}\( \phi  \){]}>\( _{P} \) =<O{[}\( \phi  \){]}>\( _{landau} \)+i\( \int  \)\( _{_{0}} \)\( ^{1} \)\( \frac{d\kappa }{\kappa ^{2}} \)
\( \int  \)D\( \phi  \) exp\{i\( \underbrace{S} \)\( ^{M}_{eff} \)+\( \varepsilon [\frac{1}{2} \)A\( _{\mu } \)A\( ^{\mu } \)-
i\( \overline{c} \)c {]}\}

~~ ~~ ~~ ~ ~ ~~ ~~ ~\( \bullet  \)\( \sum _{i}(\widetilde{\delta _{1i}[\phi } \){]}+\( \kappa  \)\( (\widetilde{\delta _{2i}[\phi } \){]}){[}\( \int d^{4}y \)
\( \overline{c} \)\( ^{\gamma } \)(y) (\( \partial  \).A\( ^{\gamma } \)){]}\( \frac{\delta ^{L}O[A,c,\kappa \overline{c}]}{\delta \underbrace{\phi _{i}}} \)\( \parallel  \)\( _{_{conn}} \)

~= <O{[}\( \phi  \){]}>\( _{landau} \)

+i\( \int  \)\( _{_{0}} \)\( ^{1} \)\( \frac{d\kappa }{\kappa ^{2}} \)<\( \sum _{i}(\widetilde{\delta _{1i}[\phi } \){]}+\( \kappa  \)\( (\widetilde{\delta _{2i}[\phi } \){]}){[}\( \int d^{4}y \)\( \overline{c} \)\( ^{\gamma } \)(y)
(\( \partial  \).A\( ^{\gamma } \)){]}\( \frac{\delta ^{L}O[A,c,\kappa \overline{c}]}{\delta \underbrace{\phi _{i}}} \)>\( \parallel _{mixed} \)\( _{_{,conn}} \)~
~~ ~ ~~~~(5.37)

We note here that in the last term, we have the connected Green's functions
in the mixed gauge with the appropriate \( \epsilon  \)-term.

\section{An Alternate and procedure for Planar Gauges}

\paragraph{In this section,we shall present an alternate ,though somewhat heuristic, procedure
that uses directly the familiar results of the section 3. This procedure applies
directly to the vacuum expectation values of gauge-invariant observables, if
not for their arbitrary Green's functions. The conclusions drawn here are however
easily \emph{directly verified} \emph{by an algebra similar to that in Section
3}; and without any need for the lack of rigor in the following derivation.We
present the former way here.}

\paragraph{To use the results of section 3, we first establish an interpolating route.
We define}

\subparagraph{\protect\( \widetilde{F}\protect \)\protect\( ^{M}\protect \){[}A{]}=\protect\( \partial .A(1-\kappa )+\kappa \eta .A\protect \)
~~ ~~ ~ ~~ ~~ ~~ ~ ~~ ~ ~~ ~~ ~~ ~~ ~~ ~ ~~ ~~ ~~ ~~ (6.1)}

\subparagraph{Then with}

S\( ^{M} \)\( _{gf} \) = -\( \frac{1}{2\lambda \eta ^{2}}\int  \) d\( ^{4}x \)
~\( \widetilde{F} \)\( ^{M} \){[}A{]}\( \partial ^{2} \)~ \( \widetilde{F} \)\( ^{M} \){[}A{]}~~
~~ ~ ~~ ~~ ~~ ~~~ ~~ ~~ ~~ ~ ~~ ~~ ~~ ~~ ~~ ~~ ~~ ~ ~~ ~(6.2)

and 

\emph{S\( ^{M} \)\( _{gh} \)= -}\( \int  \)d\( ^{4}x \) \( \overline{c} \)
\( \partial ^{2} \){[}M\( (1-\kappa )+\kappa M']c \) ~ ~~ ~~ ~~ ~ ~~ ~~ ~~
~~~ ~~ ~~ ~~ ~ ~~ ~~ ~~ ~~ ~~ ~(6.3)

and the BRS transformation

\( \delta \overline{c} \)= \{\( \widetilde{F^{M}/} \)\( \lambda \eta ^{2}\} \)
\( \delta \Lambda  \)~ ~~ ~ ~~ ~~ ~~ ~~~ ~~ ~~ ~~ ~ ~~ ~~ ~~ ~~ ~~ ~(6.4)

we have the interpolating route from the pseudo-Lorentz gauges (\( \kappa =0) \)
to the planar gauges\( (\kappa =1) \).Now, for the purpose of treating the
extra singularity introduced in the propagator on account of the \( \partial ^{2} \)
in (6.2), we interpret it as \( \partial ^{2} \)-i\( \epsilon  \) both there
and in (6.3){[}as pointed out earlier in (5.1){]}.In the momentum space,we write
the equation (6.2) as,

S\( ^{M} \)\( _{gf} \) = \( \frac{1}{2\lambda \eta ^{2}}\int  \) d\( ^{4}k \)
~\( \widetilde{F} \)\( ^{M} \){[}k{]}{[}k\( ^{2} \)+i\( \epsilon ] \)~ \( \widetilde{F} \)\( ^{M}[k] \)~

~ ~~ ~ ~=~- \( \frac{1}{2\lambda }\int  \) d\( ^{4}k \) ~\{\( \widetilde{F} \)\( ^{M} \){[}k{]}\( \sqrt{\frac{k^{2}+i\epsilon }{-\eta ^{2}}} \)\}\( ^{2} \)~~
~~ ~~~ ~~ ~~ ~~ ~ ~~ ~~ ~~ ~~ ~~ ~~ ~~ ~ ~~ ~(6.5)

where \( \widetilde{F}^{M} \){[}k{]} is the Fourier transform of \( \widetilde{F} \)\( ^{M} \){[}A{]}.We
define

F.T.\{\( \widetilde{F}^{M} \){[}k{]}\( \sqrt{\frac{k^{2}+i\epsilon }{-\eta ^{2}}} \)\}\( \equiv  \)\( \sqrt{\frac{\partial ^{2}-i\epsilon }{\eta ^{2}}} \)\( \widetilde{F} \)\( ^{M} \){[}A{]}\( \equiv  \)F\( _{M}[A] \)~~
~~ ~~ ~~ ~~ ~~ ~~ ~ ~~ ~(6.6)

We then have,

S\( ^{M} \)\( _{gf} \) = -\( \frac{1}{2\lambda }\int  \)F\( _{M}[A] \)\( ^{2} \)d\( ^{4}x \)~~
~~ ~~ ~~ ~~ ~~ ~ ~~~ ~~ ~~ ~ ~~~ ~~ ~~ ~ ~~ ~~ ~~ ~~ ~ ~~~(6.7)

and the ghost term as

\emph{S\( ^{M} \)\( _{gh} \)= -}\( \int  \)d\( ^{4}x \) \( \overline{c} \)
\( \sqrt{} \)\( \{[\partial ^{2} \)-i\( \epsilon  \) \( ]\eta ^{2}\} \)\( \frac{\delta F_{M}}{\delta A_{\mu }}D_{\mu } \)c~ 

~~~ ~~ =~\emph{-}\( \int  \)d\( ^{4}x \) \( \sqrt{} \)\( \{[\partial ^{2} \)-i\( \epsilon  \)
\( ]\eta ^{2}\} \)\( \overline{c} \) \( \frac{\delta F_{M}}{\delta A_{\mu }}D_{\mu } \)c~
~ ~~~ ~~ ~~ ~ ~~ ~~ ~~ ~~ ~ ~~~(6.8)

The last step is seen by going to the Fourier space and comparing the expressions.We
now make a linear change of variables, leading to a constant Jacobian that can
be ignored,

\( \overline{c} \)\( '=\sqrt{} \)\( \{[\partial ^{2} \)-i\( \epsilon  \)
\( ]\eta ^{2}\} \)\( \overline{c} \)~~~ ~~ ~~ ~ ~~ ~~ ~~ ~~ ~ ~~~(6.9)

Then 

\emph{S\( ^{M} \)\( _{gh} \)= -}\( \int  \)d\( ^{4}x \) \( \overline{c} \)
'\( \frac{\delta F_{M}}{\delta A_{\mu }}D_{\mu } \)c~~~ ~~ ~ ~~ ~~ ~~ ~~ ~
~ ~ ~~ ~~ ~~ ~~ ~~~~(6.10)

Now the system of gauge fixing term (6.7) and the ghost term (6.10) has been
cast in the standard FP form.We can now apply the results of the Section 3 to
interpolate between the planar and the pseudo-Lorentz gauges with

\( \Theta  \)'{[}\( \phi  \){]} =-\( i\int d^{4}y \) \( \overline{c} \)'\( ^{\gamma } \)(y)
{[}F\( _{M} \)(\( \kappa =1) \)- F\( _{M} \)(\( \kappa =0) \){]}\( ^{\gamma } \) 

~ ~~ ~~ ~~ =-\( i\int d^{4}y \) \( \overline{c} \)'\( ^{\gamma } \)(y)\( \sqrt{\frac{\partial ^{2}-i\epsilon }{\eta ^{2}}} \){[}\( \eta .A-\partial .A] \)\( ^{\gamma } \)

~ ~~ ~~ ~~= ~-\( i\int d^{4}y \) \( \overline{c} \)\( ^{\gamma } \)(y)\( [\partial ^{2}-i\epsilon ] \){[}\( \eta .A-\partial .A] \)\( ^{\gamma } \)~~
~~ ~~ ~~ ~ ~~ ~~ ~~ ~~ ~~ ~ ~~ ~~ ~~~~~ ~~ ~ ~~ ~~ ~~~(6.11)

Thus, under the field transformation

\( \phi  \)'\( _{i} \) =\( \phi  \)\( _{i} \) +\( \delta ^{L}_{iBRS}[\phi ]\Theta  \)
{[}\( \phi ] \) ~ ~~ ~~ ~~ ~ ~~ ~~ ~~ ~~~ ~~ ~~ ~~ ~ ~~ ~~ ~~ ~~~ ~~ ~~ ~~
~ ~~ ~~ ~~ (6.12)

with \( \Theta  \) given in terms of \( \Theta  \)'{[}\( \phi  \){]} of (6.11)through
relation in section 2, relates the planar gauge to the pseudo-Lorentz gauge.
Further as far as the expectation values of gauge-invariant observables are
concerned,they have been shown to have the same value in the pseudo-Lorentz
and the Lorentz gauges in Section 5.

Of course,as mentioned at the beginning of this of this section,the above heuristic
argument that allows one to make a direct use of the results of Section 3, can
be avoided and we can, with additional labor, verify the result for the field
transformation given by (6.12) directly from the Jacobian condition (2.14).We
have in fact carried out this verification.

\end{document}